\documentclass{article}
\usepackage{cite}

\usepackage[pdftex]{graphicx}
\usepackage{subcaption}
\DeclareGraphicsExtensions{.pdf,.jpg,.jpeg,.png}
\begin{document}
%
\title{Comparative Time-Series Analysis of Hip and Shoulder Rotation in Baseball Bat Swings
}
%
%
%

\author{Koga Okubo, \and
        Kanta Tachibana, 
\thanks{K. Okubo and K. Tachibana were with the Department
of System Mathematics Science, School of Informatics, Kogakuin University, Japan e-mail:jx21055@g.kogakuin.jp, kanta@cc.kogakuin.ac.jp.}}

\maketitle

\begin{abstract}
This study focuses on the rotation of the hips and shoulders during a baseball bat swing, analyzing the time-series changes in rotational angles, rotational velocities, and axes using marker position data obtained from a motion capture system with 12 infrared cameras. 
Previous studies have examined factors such as ground reaction forces, muscle activation patterns, rotational energy, angular velocity, and angles during a swing.
However, to the best of our knowledge, the hip and shoulder rotational motions have not been adequately visualized or compared. 
In particular, there is a lack of analysis regarding the coordination and timing differences between hip and shoulder movements during the swing. 
Therefore, this study aims to quantitatively compare the hip and shoulder rotational movements during the swing between skilled and unskilled players and visualizes the differences between them. 
Based on the obtained data, the study aims to improve the understanding of bat swing mechanics by visualizing the coordinated body movements during the swing.
\end{abstract}


%

\section{Introduction}
%
%
%
%


Baseball batting motions vary widely among hitters and can change even for the same hitter depending on his physical condition. 
Previous studies~\cite{Hirayama2016,Isogai2021,Horiuchi2017,Fang2021} have investigated baseball bat swings. 
Hirayama et al.~\cite{Hirayama2016} analyzed bat swing velocity and trunk rotation angles using 3D motion data, ground reaction force (GRF) using a force plate, and lower limb muscle load patterns using a musculo-skeletal model. 
They compared nine skilled players selected from a technical college baseball team with seven unskilled first-year substitute players.
Isogai et al.~\cite{Isogai2021} measured the GRF of the pivot and stance foot using a force plate.
They divided the swing motion into four phases—stance, take-back, impact, and follow-through—and compared six male baseball players from the same technical college with six male students who were not part of the baseball team.
Horiuchi et al.~\cite{Horiuchi2017} measured swing speed with 11 university baseball players as subjects.
Reflective markers were placed on the toes and heels of both feet, the tip of the bat, and the grip of the bat, and data were collected using a 3D motion analysis system.
Fang et al.~\cite{Fang2021} extracted the center of mass trajectory of the body during the swing from 2D images and compared subjects. 
However, the studies~\cite{Hirayama2016,Horiuchi2017} did not visualize or compare the kinematic time-series patterns of hip and shoulder rotation. 
In addition, the study~\cite{Isogai2021} only evaluated the maximum, minimum, and impact values of GRF in the four swing phases, without analyzing rotational movements related to body coordination. 
Furthermore, study~\cite{Fang2021} estimated data from 2D images rather than using 3D motion data.

There have also been studies of body coordination during baseball-like swing motions, such as swinging a racket, bat or club. 
The kinetic chain is critical in swing motions. 
Studies~\cite{Ishikawa2015,Zhao2019,Ozcelik2023,Ebner2020} have analyzed upper body coordination during tennis racket swing.
Ishikawa and Murakami~\cite{Ishikawa2015} studied mechanical energy and energy transfer patterns in different body segments during racket swing using inertial measurement units (IMUs) attached to seven upper body locations of three tennis players. 
Zhao et al.~\cite{Zhao2019} used motion sensors attached to the racket handle to estimate the ball speed for three types of shots—serves, ground strokes, and volleys. 
Özçelik~\cite{Ozcelik2023} compared the hit rate at the racket’s sweet spot of the racket between forehand and backhand strokes. 
However, the studies~\cite{Zhao2019,Ozcelik2023} examined only swing speed and accuracy without investigating the kinetic chain responsible for the racket movement. 
Ebner et al.~\cite{Ebner2020} analyzed 3D acceleration and angular velocity data from IMU sensors placed at the wrist, forearm, upper arm, and shoulder for eight types of tennis strokes. 

Golf swing motion has also been studied. 
Studies~\cite{Reintrakulchai2014,Kim2020,Zhang2018,Kim2017} measured acceleration and angular velocity at two locations on the upper and lower back during golf swings. 
Reintrakulchai and Kimpan~\cite{Reintrakulchai2014} used these data to compare a coach with four students. 
Kim, Zohdy, and Barker~\cite{Kim2020} analyzed amateur golfers' swings using pose estimation techniques to assess body coordination based on posture at key frames, swing tempo, and consistency. 
Zhang et al.~\cite{Zhang2018} applied machine learning to golf swing data from four skilled golfers, analyzing individual identification based on five body parts. 
Kim et al.~\cite{Kim2017} evaluated body coordination by analyzing golf swing movements from five body parts using wrist-worn IMU bands. 

The kinetic chain during cricket swings has also been studied. 
Curtis et al.~\cite{Curtis2009} used a motion capture system to record the head, both feet, and bat positions, and then applied fuzzy logic to compare body coordination between skilled and unskilled batsmen from six perspectives, including shoulder and hip movement. 
Bandara and Bačić~\cite{Bandara2020} classified batting strokes based on five coordinate points extracted from cricket batting videos. 
Gajanayake et al.~\cite{Gajanayake2021} developed a custom IMU device to measure and record bat movement in real time during swings. 
Mohonta and Sarkar~\cite{Mohonta2015} used bat-mounted inertial sensors attached to a bat to measure ten consecutive swings and analyzed the batting motion using a pendulum model to evaluate swing characteristics and performance. 

Among studies of body coordination during bat, rackets, or club swings similar to baseball batting, studies~\cite{Ebner2020,Zhang2018,Bandara2020,Gajanayake2021} focused on shoulder rotation. 
Studies~\cite{Reintrakulchai2014,Kim2020,Kim2017} examined coordination between hip and shoulder movements. 
However, none of these studies simultaneously visualized and compared the temporal patterns of hip and shoulder rotation.
Therefore, the lack of a comprehensive time-series visualization and comparison of hip-shoulder coordination in skilled and unskilled players remains an important gap in the literature.

\section{Theoretical Background}
Human body movements are driven by external forces, including GRF, gravity, and skeletal muscle contractile forces of skeletal muscles, which generate acceleration and angular acceleration in body segments. 
These forces are then converted into rotational kinetic energy by bones and joints. 
In a baseball bat swing, body segments move sequentially, transferring energy from the lower limbs to the pelvis, trunk, and upper limbs using GRF. 
This kinetic chain is essential to produce an efficient swing motion. 
Taguchi et al.~\cite{Taguchi2019} pointed out that differences in bat swings affect the load on the spine and that the timing of rotational angular velocity and acceleration is critical to maximizing bat power.

Based on these findings, bat swings are movements that are strongly influenced by both mechanical principles and physiological characteristics. 
The analysis of hip and shoulder rotational movements during a bat swing is important for improving athletic performance and preventing injuries. 
Therefore, this study focuses on hip and shoulder rotation.

\section{Method and Experiments}
\begin{figure}[htbp]
    \centering
    \includegraphics[width=0.2\linewidth]{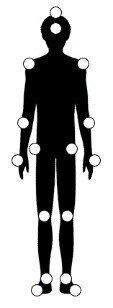}
    \caption{Positions of reflective markers.}
    \label{fig:marker_set}
\end{figure}

This study uses a motion capture system with a set of markers placed at a total of 16 locations on the body: the temples, the head, both wrists, both elbows, both shoulders, both hips, both knees, both ankles, and both toes. 
Fig.~\ref{fig:marker_set} illustrates the marker locations used for measurement.

\begin{figure}[htbp]
    \centering
    \includegraphics[width=0.8\linewidth]{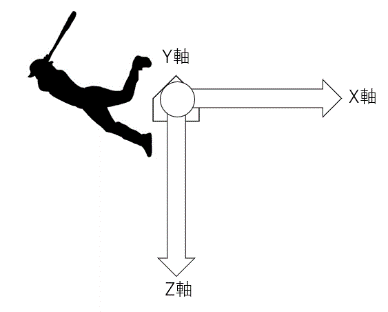} 
    \caption{Coordinate system.} 
    \label{fig:mocap_cs} 
\end{figure}

The motion capture system had a frame rate of 60 fps. 
Fig.~\ref{fig:mocap_cs} shows the coordinate system of the motion capture system.
As shown in Fig.~\ref{fig:mocap_cs}, the coordinate system is defined as follows: the $+y$ axis points vertically upward, the $+z$ axis extends from home plate toward the pitcher's mound, and the $+x$ axis follows the right-hand rule, perpendicular to both.

In this study, the swings of four subjects were measured: two registered players from the Tokyo New University Baseball League Division II (one left-handed and one right-handed) and two unregistered individuals (one left-handed and one right-handed). 
Hereafter, they are referred to as "skilled" and "unskilled" batters. 
All four subjects used a 170g plastic bat during the swings. 
During the swing, the rotational angular velocities $\omega_h (\ge 0)$ and $\omega_s (\ge 0)$ [deg/s] of hip and shoulder, respectively, are calculated for each frame. 
And the 3D rotational axis vectors of the hip $\vec{a}_h$ and the shoulder $\vec{a}_s$ are calculated, where $|\vec{a}_h|=|\vec{a}_s|=1$.

 For frame $n$, the vector $\vec{h}(n)$ is defined as from the right hip marker to the left hip marker. 
The $\omega_h$ and $\vec{a}_h$ are calculated to satisfy eq.~(\ref{eq:hip_rotation}).
Similarly, defining the vector $\vec{s}(n)$ from the right shoulder marker and the left shoulder marker, the $\omega_s$ and $\vec{a}_s$ are calculated so that they satisfy eq.~(\ref{eq:shld_rotation}).
\begin{equation}
\sin\left(\omega_h(n)\Delta t\right)\vec{a}_h(n)=\frac{\vec{h}(n)\times\vec{h}(n+1)}{|\vec{h}(n)||\vec{h}(n+1)|},
  \label{eq:hip_rotation}
\end{equation}
\begin{equation}
\sin\left(\omega_s(n)\Delta t\right)\vec{a}_s(n)=\frac{\vec{s}(n)\times\vec{s}(n+1)}{|\vec{s}(n)||\vec{s}(n+1)|},
  \label{eq:shld_rotation}
\end{equation}
where, $\times$ shows the cross product of 3D vectors and the frame interval $\Delta t=\frac{1}{60}$ [s].


\begin{figure}[htbp]
    \centering
    \begin{subfigure}{0.46\linewidth}
      \centering
      \includegraphics[width=\linewidth]{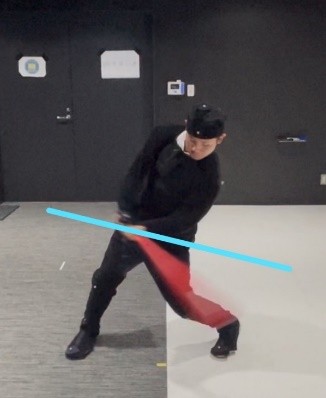} 
      \caption{Hip vector $\vec{h}(17)$}
    \end{subfigure}
    \hfill
    \begin{subfigure}{0.5\linewidth}
      \centering
      \includegraphics[width=\linewidth]{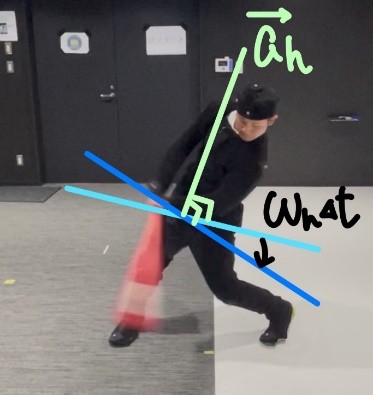}
      \caption{$\vec{h}(18)$, $\vec{a}_h(17)$ and $\omega_h(17)$}
    \end{subfigure}
    \caption{Hip rotation speed and axis from frame 17 to frame 18.}
    \label{fig:illustration_hip_rot}
\end{figure}

The inclination angle $\varphi_h(n)$ of the hip’s plane of rotation relative to the horizontal plane is calculated for frame $n$.
$$\cos\left(\varphi_h(n)\right)=\vec{e}_y\cdot\vec{a}_h(n),$$
$$\cos\left(\varphi_s(n)\right)=\vec{e}_y\cdot\vec{a}_s(n),$$
where $\vec{e}_y$ is the unit vector pointing vertically downward for left hitter and upward for right hitter.
The time-series graphs of $\varphi_h(n)$ and $\varphi_s(n)$ are compared between skilled and unskilled players. 
The similarities and differences are analyzed and discussed.

\section{Results}
The results of the experiments on the rotation angle, rotation angular velocity, and axis of rotation axis are described. 
This experiment was conducted with four participants, two each with and two without baseball experience.

\subsection{Horizontal Rotation Angle}
\begin{figure}[htbp]
    \centering
    \includegraphics[width=0.5\linewidth]{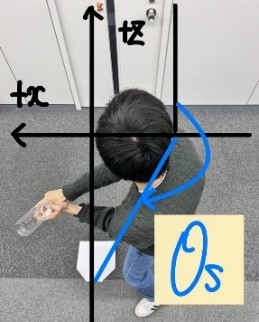} 
    \caption{Definition of shoulder rotation angle $\theta_s$ in the horizontal plane.} 
    \label{fig:illustration_horizontal_rot} 
\end{figure}

Fig.~\ref{fig:illustration_horizontal_rot} defines the shoulder rotation angle $\theta_s(n)$ [deg] in the horizontal plane.
Similarly, the hip rotation angle $\theta_h(n)$ is defined with the $+z$-direction as 0 [deg] and is measured clockwise when viewed from above. 
In Fig.~\ref{fig:illustration_horizontal_rot}, the angle between the vector $\vec{s}(n)$ projected onto the horizontal plane and the $+z$-direction is shown blue.

\begin{figure}[htbp]
    \centering
    \includegraphics[width=\linewidth]{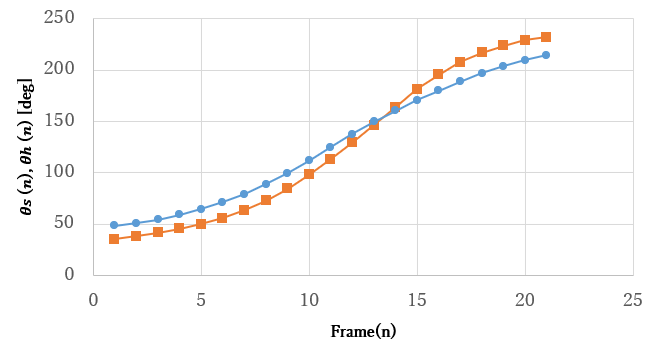} 
    \caption{Horizontal rotation angle of hip (blue) and shoulder (orange) in the skilled left batter.} 
    \label{fig:rot_angle_result_skilled_left} 
\end{figure}

\begin{figure}[htbp]
    \centering
    \includegraphics[width=\linewidth]{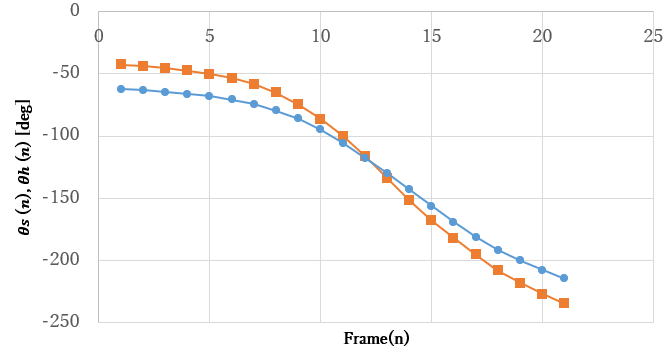} 
    \caption{Horizontal rotation angle of hip (blue) and shoulder (orange) in the skilled right batter.} 
    \label{fig:rot_angle_result_skilled_right} 
\end{figure}

\begin{figure}[htbp]
    \centering
    \includegraphics[width=\linewidth]{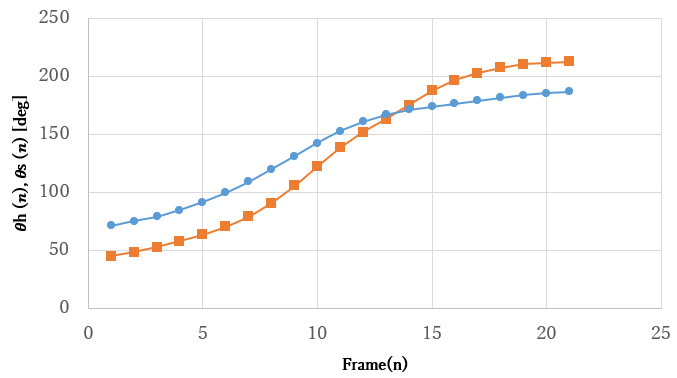} 
    \caption{Horizontal rotation angle of hip (blue) and shoulder (orange) in the unskilled left batter.} 
    \label{fig:rot_angle_result_unskilled_left} 
\end{figure}

\begin{figure}[htbp]
    \centering
    \includegraphics[width=\linewidth]{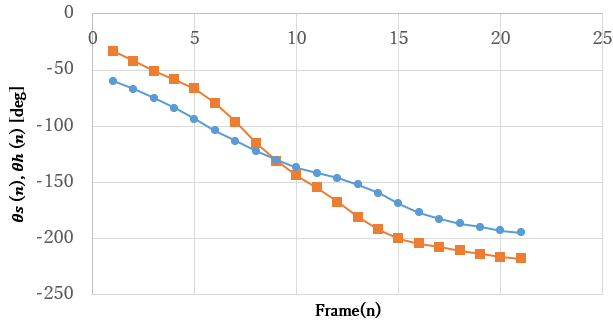} 
    \caption{Horizontal rotation angle of hip (blue) and shoulder (orange) in the unskilled right batter.} 
    \label{fig:rot_angle_result_unskilled_right} 
\end{figure}

The beginning frame of batting motion is defined as that at which either of the hip or shoulder rotation starts, i.e., the Y-component of the rotation axis vector becomes positive for right-handed batters and negative for left-handed batters. 
Also, the end frame of batting motion is defined that the last frame of the rotation continues for either hip or shoulder. 
Figs.~\ref{fig:rot_angle_result_skilled_left} and \ref{fig:rot_angle_result_skilled_right} show the hip and shoulder rotation angles in the horizontal plane for skilled players. 
The horizontal axis represents the frame. 
The vertical axis represents the angle in the horizontal plane. 
Figs.~\ref{fig:rot_angle_result_unskilled_left} and ~\ref{fig:rot_angle_result_unskilled_right} show the hip and shoulder rotation angles for unskilled players.



\subsection{Rotation Speed}
\begin{figure}[htbp]
    \centering
    \includegraphics[width=\linewidth]{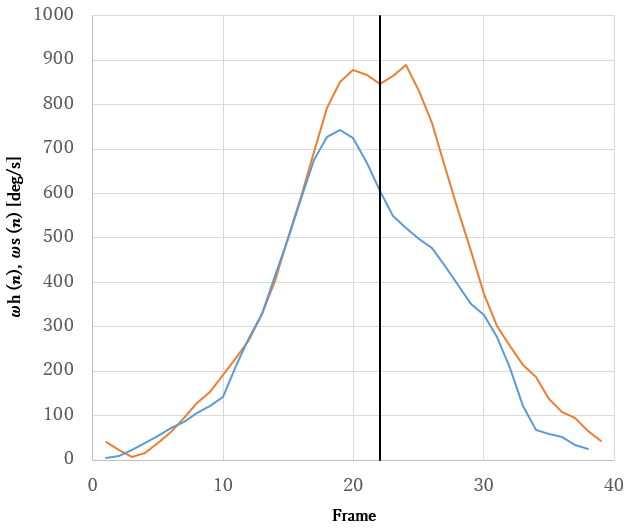} 
    \caption{Rotation speed of hip (blue) and shoulder (orange) in the skilled left batter.} 
    \label{fig:rot_speed_result_skilled_left} 
\end{figure}

\begin{figure}[htbp]
    \centering
    \includegraphics[width=\linewidth]{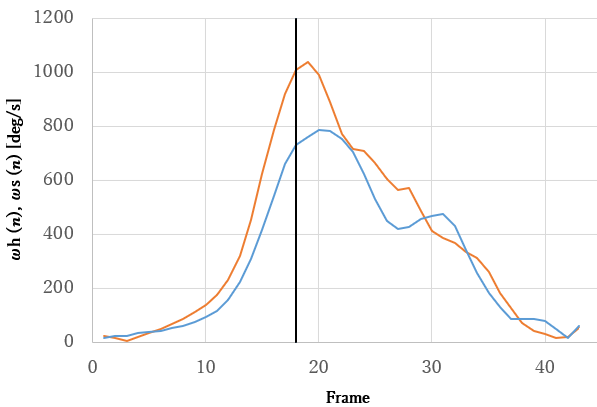} 
    \caption{Rotation speed of hip (blue) and shoulder (orange) in the skilled right batter.} 
    \label{fig:rot_speed_result_skilled_right} 
\end{figure}

\begin{figure}[htbp]
    \centering
    \includegraphics[width=\linewidth]{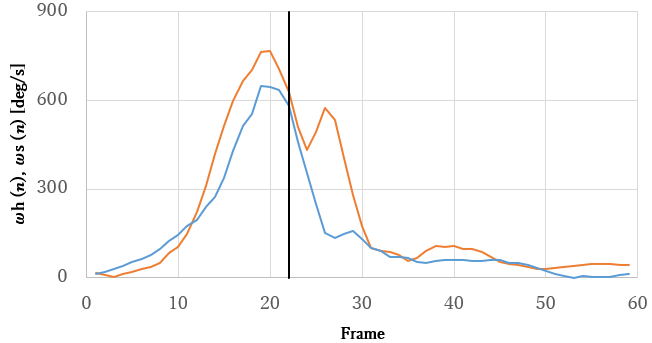} 
    \caption{Rotation speed of hip (blue) and shoulder (orange) in the unskilled left batter.} 
    \label{fig:rot_speed_result_unskilled_left} 
\end{figure}

\begin{figure}[htbp]
    \centering
    \includegraphics[width=\linewidth]{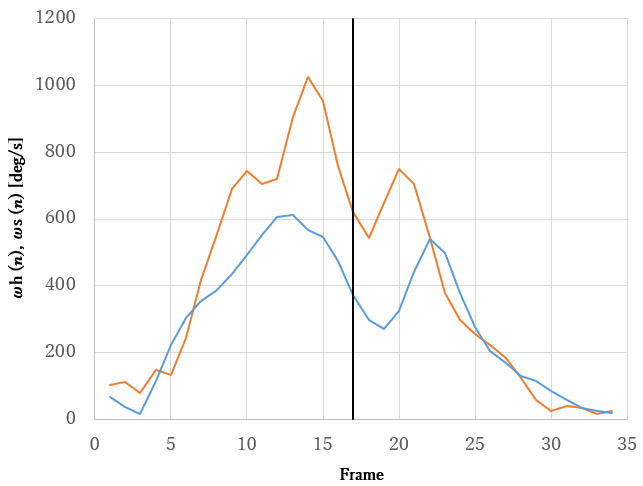} 
    \caption{Rotation speed of hip (blue) and shoulder (orange) in the unskilled right batter.} 
    \label{fig:rot_speed_result_unskilled_right} 
\end{figure}

Figs.~\ref{fig:rot_speed_result_skilled_left} and \ref{fig:rot_speed_result_skilled_right} show the hip and shoulder rotation speed during the swing for skilled players. 
In these figures, the horizontal axis represents the frame number. 
The vertical axis represents the rotation speed [deg/s]. 
The blue line represents the hip rotation speed, and the orange line represents the shoulder rotation speed.
The black vertical line marks the frame just before the wrist snap during the swing.
Similarly, Figs.~\ref{fig:rot_speed_result_unskilled_left} and \ref{fig:rot_speed_result_unskilled_right} show the rotation speed for unskilled players.



\subsection{Rotation Axis}
\begin{figure}[htbp]
    \centering
    \includegraphics[width=\linewidth]{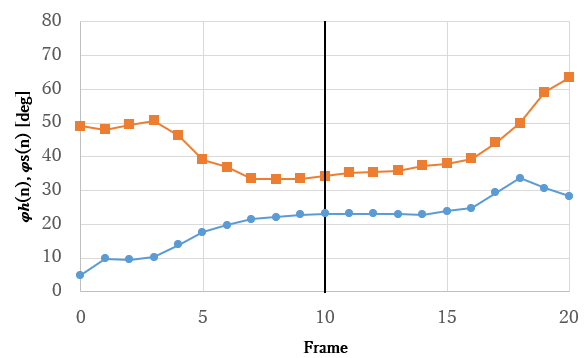} 
    \caption{Inclination of rotation axis of hip (blue) and shoulder (orange) in the skilled left batter.} 
    \label{fig:axis_result_skilled_left} 
\end{figure}

\begin{figure}[htbp]
    \centering
    \includegraphics[width=\linewidth]{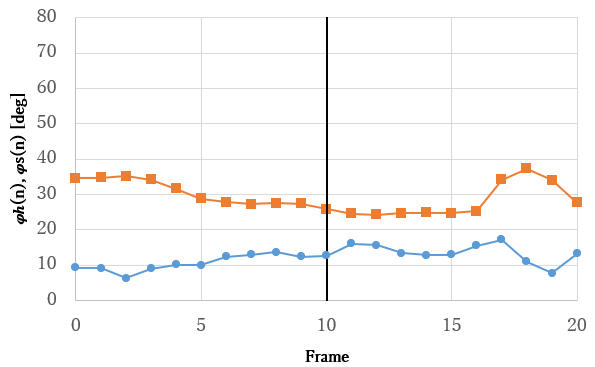} 
    \caption{Inclination of rotation axis of hip (blue) and shoulder (orange) in the skilled right batter.} 
    \label{fig:axis_result_skilled_right} 
\end{figure}

\begin{figure}[htbp]
    \centering
    \includegraphics[width=\linewidth]{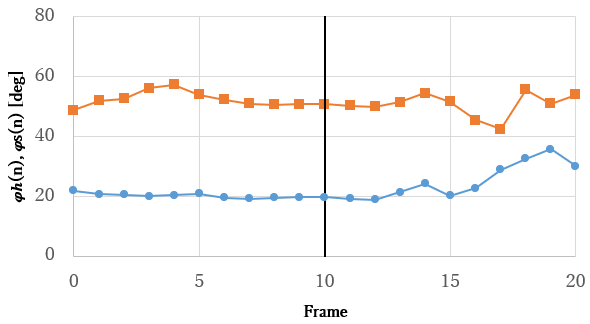} 
    \caption{Inclination of rotation axis of hip (blue) and shoulder (orange) in the unskilled left batter.} 
    \label{fig:axis_result_unskilled_left} 
\end{figure}

\begin{figure}[htbp]
    \centering
    \includegraphics[width=\linewidth]{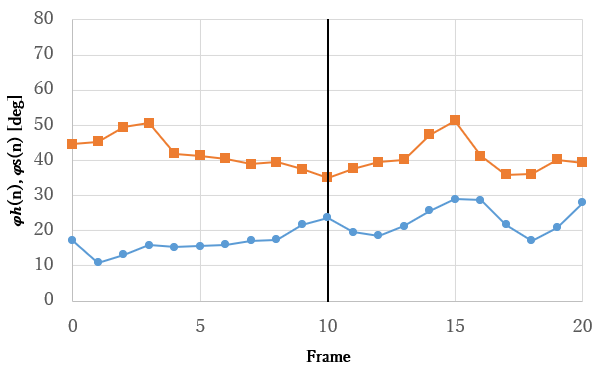} 
    \caption{Inclination of rotation axis of hip (blue) and shoulder (orange) in the unskilled right batter.} 
    \label{fig:axis_result_unskilled_right} 
\end{figure}

Figs.~\ref{fig:axis_result_skilled_left} and \ref{fig:axis_result_skilled_right} show the angle [deg] between the rotation plane and the horizontal plane for the skilled players.
The 10 frames before and 10 frame after the wrist snap frame, a total of 21 frames are shown. 
In these figures, the blue line represents the inclination of hip rotation plane $\theta_h$. 
The orange line represents that of shoulder rotation plane $\theta_s$.
The black vertical line marks the frame just before the wrist snap during the swing. Similarly, Figs.~\ref{fig:axis_result_unskilled_left} and \ref{fig:axis_result_unskilled_right} show the tilt of the rotation planes for unskilled players.



\section{Discussions}
The differences between skilled and unskilled players in terms of hip and shoulder rotation during the batting motion are outlined below.
\begin{enumerate}
    \renewcommand{\labelenumi}{\Alph{enumi})} 
    \item \emph{Wrist snap timing and maximum shoulder rotation speed}
    \item \emph{Stability of the hip and shoulder rotation axis}
\end{enumerate}

\subsection{Wrist snap timing and maximum shoulder rotation speed}
    For the skilled players, the shoulder rotation speed peaks at or just after the moment of the wrist snap. 
    Figs.~\ref{fig:rot_speed_result_skilled_left} and \ref{fig:rot_speed_result_skilled_right} show that the peaks of both skilled players' shoulder rotation are 1 frame after the wrist snap.
    In contrast, for unskilled players, the shoulder rotation speed begins to slow down at the moment of the wrist snap.
    Figs.~\ref{fig:rot_speed_result_unskilled_left} and \ref{fig:rot_speed_result_unskilled_right} show that the peaks of both unskilled players' shoulder rotation are 3 frame before the wrist snap.

\subsection{Stability of the hip and shoulder rotation axis}
    Skilled players maintain a stable rotation axis during the 11 frames surrounding the impact. 
    In contrast, unskilled players show a greater tilt in the shoulder rotation axis, and both the hip and shoulder rotation axes show significant variability. 
    Table~\ref{table:axis_stability} summarizes the angles of the hip and shoulder rotation planes relative to the horizontal plane (mean $\pm$ standard deviation) for the 5 frames before and 5 frames after, totally 11 frames around the wrist snap.
    \begin{table}[htbp]
        \centering
        \caption{Mean $\pm$ standard deviation of rotation plane tilt.}
        \label{table:axis_stability}
        \begin{tabular}{c|cc}
             & $\varphi_h$[deg] &$\varphi_s$[deg] \\
            \hline
            skilled, left & 22.1$\pm$ 1.8 & 35.7$\pm$2.0\\
            skilled, right & 13.2$\pm$ 1.6 & 26.1$\pm$1.6\\
            \hline
            unskilled, left & 20.1$\pm$ 1.5 & 51.6$\pm$1.5\\
            unskilled, right & 20.5$\pm$ 4.3 & 40.8$\pm$4.6\\
        \end{tabular}
    \end{table}

\begin{figure}[htbp]
    \centering
    \includegraphics[width=0.6\linewidth]{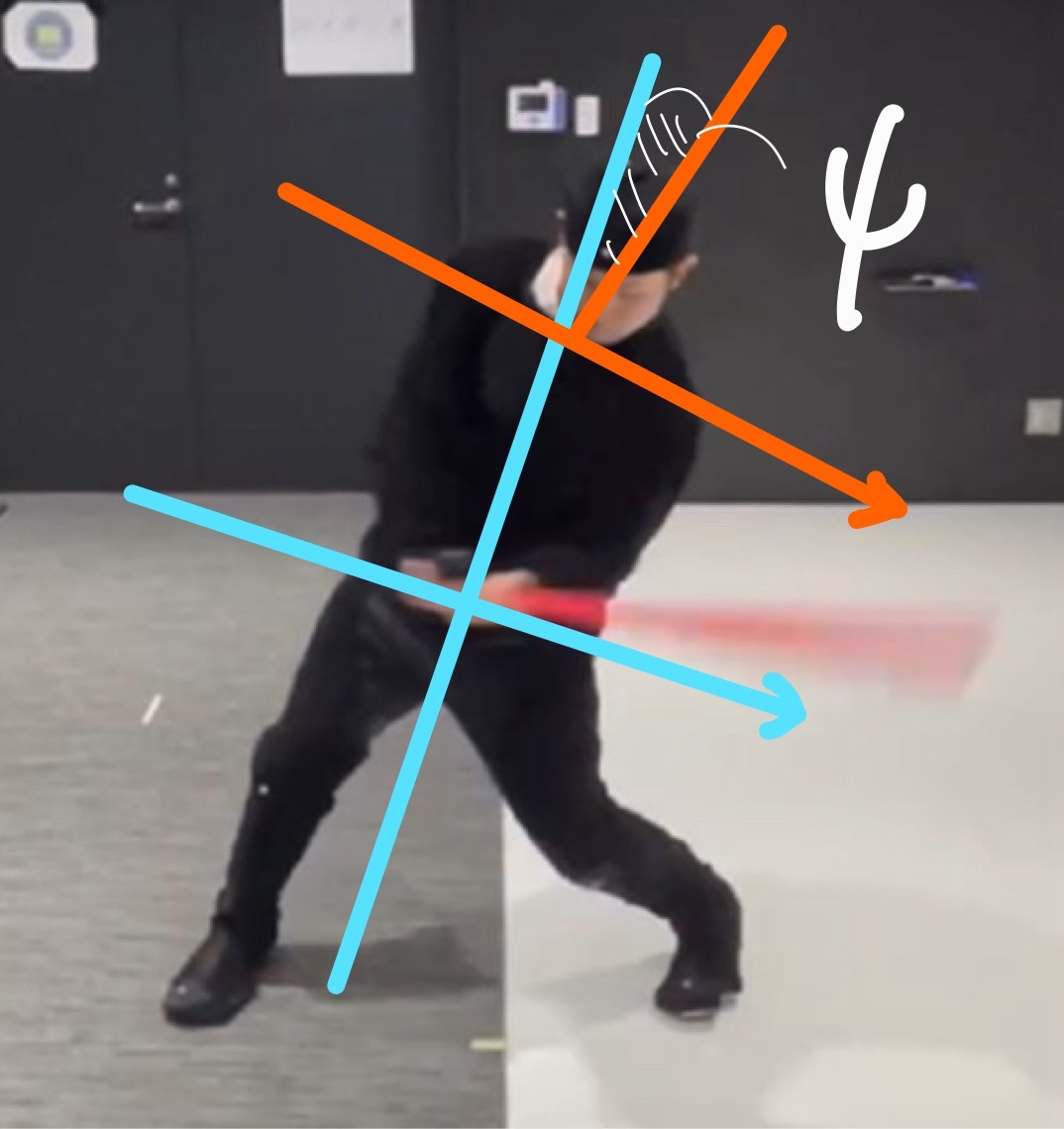} 
    \caption{Relative angle $\psi$ between hip and shoulder rotation axes.} 
    \label{fig:illustration_relative_angle} 
\end{figure}

\begin{figure}[htbp]
    \centering
    \includegraphics[width=\linewidth]{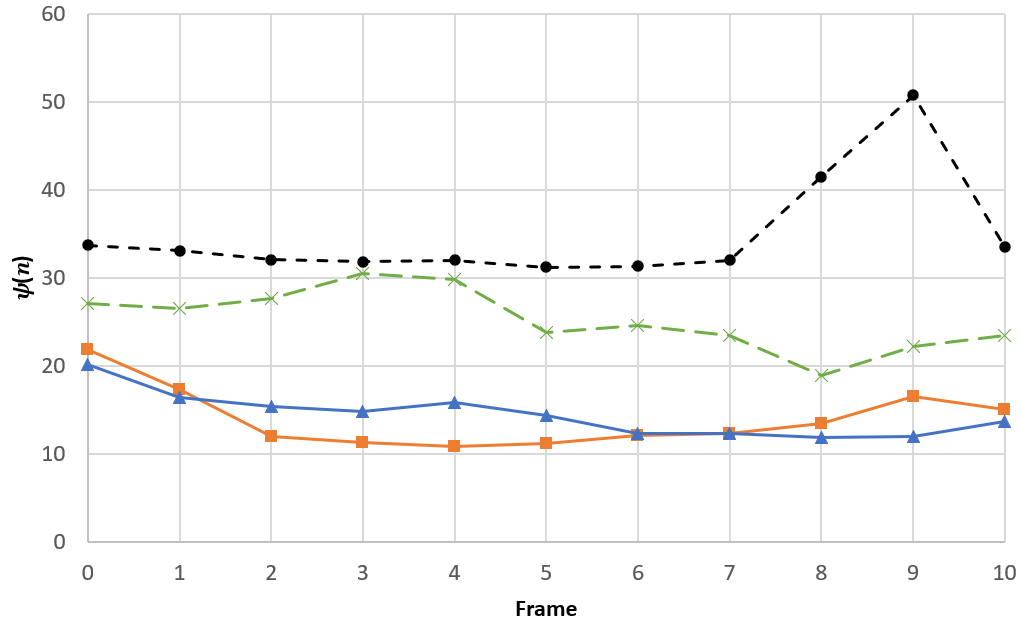} 
    \caption{Relative angle $\psi$ of rotation axes for skilled and unskilled players.} 
    \label{fig:relative_angle_result} 
\end{figure}

With the relative angle $\psi(n)$ between the rotation planes of hip and shoulder for frame $n$, the following equation is satisfied,
$$\cos(\psi(n))=\vec{a}_s(n)\cdot \vec{a}_h(n).$$
Fig.~\ref{fig:illustration_relative_angle} illustrates the relative angle $\psi(n)$. Fig.~\ref{fig:relative_angle_result} shows the relative angles of the four players. 
Frame 5 represents the wrist snap of each player. 
Skilled players are shown as straight lines and unskilled players are shown as dashed lines.

\section{Conclusion}
The purpose of this study was to analyze time-series patterns of the rotational axes and angles of the waist and shoulders during a batter's swing. 
A motion capture system was used to collect swing data and to visualize the kinetic chain of the most important rotational movements, hip and shoulder rotation.

The analysis revealed significant differences between skilled and unskilled participants, particularly in (A) shoulder rotational velocity at the moment of wrist rotation and (B) stability of the waist and shoulder rotational axes. 
In addition, the experiment revealed individual differences in the use of the hip and shoulders. 
By visualizing the coordinated movement of the body during swing and comparing swing movements, this study provided insight into swing mechanics.

For further improvement, future challenges include increasing the number of participants to ensure statistically significant differences by comparing two groups, skilled and unskilled players.
The population size of skilled players is approximately 30,000 including NPB, Nippon Professional Baseball, and college league baseball players.
The general population of the same age group in Japan as unskilled players is approximately 10 million.
Considering these population sizes, statistical significance is achieved with a sample size of 12 to 30 participants per group.
Also, increasing the number of swing trials will contribute to investigate both inter- and intra-individual variation.


%








\bibliographystyle{IEEEtran}
\bibliography{./refs.bib}
%

%








\end{document}